\documentclass[conference,letterpaper]{IEEEtran}

\usepackage[dvipsnames]{xcolor}
\usepackage{cite}
\usepackage{amsmath,amssymb,amsfonts}
\usepackage{algorithmic}
\usepackage{graphicx}
\usepackage{textcomp}
\usepackage{xcolor}
\usepackage{fancyvrb}
\usepackage{multirow}
\usepackage{url}
\usepackage{xspace}
\usepackage{booktabs}
\usepackage{tabularx}
\usepackage{verbatim}
\usepackage{tikz}
\usepackage{pgfplots}
\usepackage{xurl}
\usepackage{hyperref} 
\usepackage{makecell}

\DefineVerbatimEnvironment{comentario}{Verbatim}{fontsize=\footnotesize}

\usepackage{listings}
\definecolor{codegray}{gray}{0.95}
\lstdefinestyle{pythonstyle}{
    language=Python,
    basicstyle=\ttfamily\small\color{black},
    keywordstyle=\color{blue},
    commentstyle=\color{green!50!black},
    stringstyle=\color{orange},
    numbers=left,
    numberstyle=\tiny\color{black},
    stepnumber=1,
    numbersep=5pt,
    showspaces=false,
    showstringspaces=false,
    showtabs=false,
    breaklines=true,
    tabsize=4,
    captionpos=b
}

\def\BibTeX{{\rm B\kern-.05em{\sc i\kern-.025em b}\kern-.08em
    T\kern-.1667em\lower.7ex\hbox{E}\kern-.125emX}}
\begin{document}

\title{Evaluation of Domain-Specific Architectures for General-Purpose Applications in Apple Silicon}

\author{\IEEEauthorblockN{Álvaro Corrochano López}
\IEEEauthorblockA{\textit{Computer Science Faculty} \\
\textit{Complutense University of Madrid}\\
Madrid, Spain \\
acorroch@ucm.es \\
0009-0000-8160-7194}
\and
\IEEEauthorblockN{Carlos García Sánchez}
\IEEEauthorblockA{\textit{Computer Science Faculty} \\
\textit{Complutense University of Madrid}\\
Madrid, Spain \\
garsanca@ucm.es\\
0000-0002-3470-1097}

}

\maketitle

\begin{abstract}
The rise of AI and its growing computational demands have driven the integration of domain-specific accelerators (such as GPUs, TPUs, and NPUs) across the entire computing infrastructure. Following the precedent set by the GPGPU which popularized GPUs for general-purpose tasks, this research asks whether this phenomenon can be replicated with specialized accelerators like NPUs in new contexts. This paper evaluates the potential of the Apple Neural Engine (ANE) designed for high energy efficiency in Machine Learning workloads, in the context of general-purpose HPC applications. We evaluate the performance and energy consumption of classic HPC algorithms such as GEMM, Jacobi or Multigrid methods on Apple's ANE across the M1 and the latest M4 architectures. Results confirm that, when algorithms are properly adapted, the ANE achieves competitive performance (up to 3.8 TFlops on the M4-Pro, comparable to the GPU's 4.7 TFlops on the same SoC for GEMM operation) while demonstrating significantly superior energy efficiency (e.g., GEMM consumes 5.2 Watts on the ANE versus 24 Watts on GPU counterpart in M4 architectures).
\end{abstract}

\begin{IEEEkeywords}
Hardware acceleration, ANE, NUC, Evaluation, HPC, Apple, Jacobi, GEMM, Multigrid
\end{IEEEkeywords}

\section{Introduction}

With the advent of the Exascale era, we have witnessed an explosion of workloads driven by the rise of Artificial Intelligence (AI) and the processing of massive amounts of data spurred by the Big Data era. This trend, combined with the increasing computing power of High-Performance Computing (HPC) systems, has propelled the creation of more sophisticated techniques which the most notable example is the recent Generative AI. Driven by advancements in technology and system architecture, HPC nodes integrate not only a growing number of high-end processors but also domain specific accelerators. As an example, the use of GPUs for general-purpose applications is the last decades facilitated their popularization~\cite{Opt_for_GPUs} across different fields of science and boosted their use. Moreover, the widespread availability and advancement of GPUs have fueled the recent AI revolution~\cite{AI_accelerator}, allowing for the training and deployment of ever-more complex models.

One of the key milestones in the history of AI was the emergence of AlexNet~\cite{Alexnet}, a precursor to the development of modern models for image classification and recognition, natural language processing, and Generative AI. The escalating computational demand of these applications has driven the design of specialized accelerators—such as TPUs (Tensor Processing Units) and NPUs (Neural Processing Units)-which are now integrated across the entire computing stack~\cite{DL_HW_accelerator}, from embedded systems and desktop computers to high-performance servers, HPC infrastructures, and large-scale data centers.

NPUs~\cite{9739030,reed2022reinventinghighperformancecomputing} have established themselves as a fundamental component of desktop computing, especially since the introduction of Transformers~\cite{vaswani2023attentionneed}, which power widely used tools such as ChatGPT and DeepSeek. This increasing interest is reflected in the development of dedicated NPU hardware by leading technology companies, including AMD\footnote{\url{https://www.amd.com/es/products/processors/consumer/ryzen-ai.html}}, Samsung\footnote{\url{https://semiconductor.samsung.com/support/tools-resources/dictionary/the-neural-processing-unit-npu-a-brainy-next-generation-semiconductor/}}, Intel\footnote{\url{https://www.intel.la/content/www/xl/es/products/docs/accelerator-engines/ai-engines.html}}, and Apple\footnote{\url{https://machinelearning.apple.com/research/neural-engine-transformers}}. 

A clear example of this trend was the launch of Apple's M architectures in 2020 with the M1 chip~\cite{AppleM1-MPR}, based on a System on Chip (SoC) which integrated high-performance and high-efficiency multi-core CPU, a GPU, a matrix-multiplication co-processor, and dedicated Neural Engine (ANE). This integration lets devices use their power efficiently, delivering high performance for everything from daily tasks to heavy graphics processing and AI applications. Since then, Apple has evolved the M1 into more modern versions, subsequently releasing the M2, M3, and M4 chips, with the M5 expected to launch in 2026.

Just as GPUs gained popularity in non-graphics contexts through the past GPGPU (General-Purpose computing on Graphics Processing Units) movement, we do not rule out this phenomenon recurring for other specialized accelerators like NPUs (Neural Processing Units). This work evaluates the use of Apple's ANE and the other accelerators in a general-purpose application context to determine if this type of devices can be competitive-not only in terms of performance but also energy efficiency-compared to other devices currently available in a desktop system. As related work that explore NPUs for other propuses we can find approaches such as the Parrot Transformation~\cite{6493641} explore the replacement of imperative code segments with neural networks to improve program performance and energy efficiency. Other example is the research of Hubner et al.~\cite{hubner2025} which provide an in-depth analysis of the Apple Silicon architecture, highlighting its main computing units: the CPU, GPU, and ANE. Apart from these papers, the ANE remains relatively underexplored, although some studies have begun to examine it. 

The remainder of the paper is structured as follows: Section~\ref{AppleSilicon} and~\ref{experimental_enviroment} introduce the Apple Silicon architecture and environment for developping, Section~\ref{target_apps} presents the applications used in this study, Section~\ref{results} discussed performance results achieved, and finally Section~\ref{conclusions} provides concluding remarks and potential future research directions.

\section{Apple Silicon}
\label{AppleSilicon}
\subsection{CPU}
The Apple Silicon CPU adopts a \textit{big.LITTLE} architecture, combining two types of cores: high-performance cores and energy-efficient cores. This design allows workloads to be distributed across different cores with the dual goal of enhancing performance and optimizing power consumption. The CPU supports programming in C, C++, Objective-C, and Swift. \\
In addition, the CPU provides support for vector operations through 128-bit NEON instructions and relies on a specialized coprocessor for matrix computations: the Apple Matrix eXtension (AMX).~\cite{hubner2025}

\subsubsection{AMX}
The AMX is a coprocessor designed to accelerate matrix operations in conjunction with the CPU. A significant enhancement was introduced in the latest Apple Silicon generation (M4), which added support for the ARM Scalable Matrix Extension (SME), an ARM instruction set tailored for matrix computations. From a programming perspective, AMX can be accessed through the \texttt{Accelerate} library\footnote{\url{https://developer.apple.com/documentation/accelerate}}, a high-level C++ framework for numerical computations\footnote{\url{https://zhen8838.github.io/2024/04/23/mac-amx_en/}}. AMX can also be leveraged indirectly via CoreML; however, in this case, the execution flow is abstracted, and the framework autonomously decides whether computations are performed on the CPU or offloaded to AMX. At present, there is no official documentation available for programming AMX at a low level.

\subsection{GPU}
Series M GPU uses tile-based deferred rendering (TBDR) to divide a scene into tiles and render it tile by tile. This device can be programmed with Metal API,\footnote{\url{https://developer.apple.com/metal/}} a low-level programming framework that can be used with Metal Shading Language (MSL).\footnote{\url{https://developer.apple.com/metal/Metal-Shading-Language-Specification.pdf}} Lastly, for high-level programming, Metal Performance Shaders (MPS) and CoreML frameworks\footnote{\url{https://developer.apple.com/documentation/coreml}} can be used.

\subsection{ANE}
The Apple Neural Engine (ANE) is a hardware accelerator~\cite{hubner2025} specifically optimized for machine learning (ML) workloads. Unlike the AMX, the ANE operates as a standalone processing unit rather than as a coprocessor. Firstly released as part of the A11 chip found in iPhone X and also incorporated in Apple Silicon devices. It is optimized to handle FP16 data exclusively\footnote{\url{https://apple.github.io/coremltools/docs-guides/source/typed-execution.html}}, although official technical documentation remains scarce apart from the US Patent.\footnote{https://patents.google.com/patent/US9836691B1/de} Programming the ANE is only possible through the CoreML framework; however, developers have no direct control over its configuration. 

\subsection{CoreMLTools}
CoreMLTools is Apple’s Python framework for machine learning operations. It is important to note that the election of this framework has the advantage of portability, allowing the execution of any application among Apple's SoC devices, such as CPU, GPU, or ANE. However, it does not permit direct control of each device individually. In practice, the CPU may still be used in all execution modes if CoreML deems it optimal or if an operation or layer is not supported by GPU/ANE. The available execution options are:
\begin{itemize}     
     \item \textit{ct.ComputeUnit.ALL}: Execution distributed among CPU, GPU, and ANE.
     \item \textit{ct.ComputeUnit.CPU\_ONLY}: Execution restricted to CPU.
     \item \textit{ct.ComputeUnit.CPU\_AND\_GPU}: Execution distributed between CPU and GPU.
     \item \textit{ct.ComputeUnit.CPU\_AND\_ANE}: Execution distributed between CPU and ANE.
\end{itemize}

Core ML Tools facilitate the integration of models from popular third-party frameworks such as TensorFlow and PyTorch by transforming them into the native Core ML format (.mlmodel or .mlpackage). However, there exists an important variability in supported data types across different hardware. While both 32-bit floating-point (fp32) and 16-bit floating-point (fp16) formats are supported by the CPU and GPU, the ANE is restricted to the fp16 format for model inference.\footnote{Data types supported: \url{https://apple.github.io/coremltools/docs-guides/_images/ml-program-runtime.png}}

\section{Experimental Enviroment}
\label{experimental_enviroment}
Experiments were conducted on two different Mac systems to compare performance and observe the evolution of Apple hardware: a Mac Mini M1 and a Mac Mini M4 Pro. These devices were selected because the Mac Mini M1 represents the first generation of Apple’s System-on-Chip (SoC) for Mac devices, while the M4 Pro was the latest model available at the time of this study. Using the first and most recent Apple Silicon chip models provides a broad perspective on the performance evolution of Apple Silicon. This allows not only for comparison among the three integrated processing units on the chip but also for evaluating the improvement of each component across generations. \\
The M1 was a landmark product for Apple, as it was the first computer in the Apple Silicon series—the SoC family that Apple currently markets. The M1 (2020) includes a 16-core Neural Processing Unit (NPU), an 8-core GPU, and an 8-core CPU based on the ARM big.LITTLE architecture, featuring four efficiency cores and four performance cores. This model also integrates an AMX coprocessor that works alongside the CPU. The M4 Pro (2024) shows significant advancements compared to the first Apple Silicon model. It features a 12-core CPU with eight performance cores and four efficiency cores. The GPU doubles the number of cores to sixteen, and the NPU maintains the same core count. Tabl~ \ref{tab:architectures} summarizes the specifications of both computers.

\begin{table*}[htbp]
\small
\caption{Specifications of M1 and M4 Pro platforms.}
\label{tab:architectures}
\centering
\resizebox{\linewidth}{!}{%
\begin{tabular}{cccccccccccccc}
\multirow{2}{*}{Arch.}           & \multicolumn{2}{c}{\textbf{CPU}} & \multicolumn{2}{c}{\textbf{P-cores}} & \multicolumn{2}{c}{\textbf{E-cores}}  & {\textbf{RAM}} &\multicolumn{2}{c}{\textbf{GPU}} & {\textbf{NPU}}& {\textbf{Trans.}} & {\textbf{Bwith.}}\\

                                 & P-Cores   & E-cores              & L1               & L2        & L1               & L2        & \multicolumn{1}{c}{}   & Cores           & ALU            &       &   & \\
\textbf{M1}                      & 4   (3.2 GHz)   & 4 (2.1 GHz)                    & 128KB            & 12 MB             & 64 KB    & 4 MB     & 16 GB    & 8    & 1024     & 11 TOPS & 16 billion                  & 68 GB/s  \\
\textbf{M4 Pro} & 4 + 4 (4.51 GHz) & 4 (2.59 GHz) & 128 KB & 16 + 16 MB & 64 KB & 4 MB & 24 GB & 16 & 2048 & 38 TOPS & N. A. & 273 GB/s \\                
\end{tabular}
}
\end{table*}

Both systems were programmed in Python using CoreML, as this framework provides access to all three processing units (CPU, GPU, and ANE). The libraries and frameworks version used are summarized below:
\begin{itemize}
    \item \textbf{Python:} chosen version 3.9.6 as the main programming language due to CoreMLTools support.
    \item \textbf{PyTorch:} employed for model creation prior to conversion into the CoreML format. Version 2.3.0 was used.
    \item \textbf{Torchvision:} Used for handling model architectures and datasets. Version 0.18.0 was used.
    \item \textbf{NumPy:} version 2.0.2 for matrix operations.
    \item \textbf{CoreMLTools:} version 8.0b2 framework for machine learning operations.
\end{itemize}

In addition, energy consumption measurements required the use of both \texttt{powermetrics}\footnote{\url{https://ss64.com/mac/powermetrics.html}} and, initially, \texttt{asitop} tools\footnote{\url{https://github.com/tlkh/asitop}}. \texttt{powermetrics} is a macOS command-line utility that provides detailed information on the system's energy usage. For this study, the sampling for power consumption has been established to 100 ms. Finally, Xcode\footnote{\url{https://developer.apple.com/xcode/}} was employed to analyze operator compatibility with the ANE by inspecting models directly on its interface. The code developed in this work can be found in a public repository. \footnote{\url{https://github.com/Corrochano/Evaluation-of-Domain-Specific-Architectures-for-General-Purpose-Applications-in-Apple-Silicon}}.

\section{Target Applications}
\label{target_apps}
This work is divided into two main stages: (1) Evaluation of Apple Silicon Devices for AI on the CPU, GPU, and ANE using a widely adopted AI model, specifically YOLO (versions v3 and v11), and (2) General-Purpose Algorithm Testing extending the experiments to HPC context through the used of benchamrks as General Matrix Multiply (GEMM), Jacobi, and Multigrid. For this task two algorithms were implemented to solve the heat equation, which models the diffusion of heat over time within a discretized grid (see the Equation~\ref{eq:heat-equation}).

\begin{equation}
\frac{\partial u}{\partial t} = \left( \frac{\partial^2 u}{\partial x_1^2} + \frac{\partial^2 u}{\partial x_2^2} + \dots + \frac{\partial^2 u}{\partial x_n^2} \right)
\label{eq:heat-equation}
\end{equation}

The following subsections describes the code migration to CoreML.

\subsection{YOLOv3}
YOLOv3 (2018) is an object detection model capable of classifying 80 different categories. This model was selected to evaluate the performance of CPU, GPU and ANE in the context of AI workload. The model was obtained from Apple’s official CoreML examples repository.\footnote{\url{https://developer.apple.com/machine-learning/models/}} The experiments were executed 100 times to collect realistic performance measurements.

\subsection{YOLOv11}
YOLOv11 (2024) is one of the most recent versions of the YOLO family developed by Ultralytics. The models were obtained from the official GitHub repository. Since the provided implementations are in PyTorch, convertion into the CoreML \texttt{.mlmodel} format is mandatory (Apple’s official conversion guidelines\footnote{\url{https://apple.github.io/coremltools/docs-guides/source/convert-to-ml-program.html}}). This model was also chosen because YOLOv3 is now relatively outdated. YOLOv11 provides a more up-to-date benchmark aligned with the current state of AI model development. 

\subsection{GEMM}
GEMM was the first general-purpose algorithm evaluated in this study. This algorithm was selected due to its central role in the field of High-Performance Computing (HPC) and its widespread use in scientific and industrial applications, in addition to its use on a huge number of AI operations. For that last use, GEMM is natively implemented in PyTorch, which facilitated its integration into our experimental framework.



GEMM was readily adapted for execution on the ANE by utilizing the PyTorch operation \textit{torch.matmul(A, B)}\footnote{\url{https://docs.pytorch.org/docs/stable/generated/torch.matmul.html}} within the forward method of the model.

\subsection{Jacobi}
\label{ssect:jacobi}
The Jacobi method is an iterative algorithm used to solve systems of differential equations. Particularly, the heat equation described in Equation~\ref{eq:heat-equation} has been selected as benchmark. Unlike GEMM, adapting Jacobi for execution on the ANE required additional modifications that are outlined below:

\begin{itemize}     
     \item Start the grid with the initial temperature of each point.
     \item On each iteration, calculate the new temperature of each point using Stencil-5.
     \item Repeat this process several times or the convergence criterion is achieved.
\end{itemize}

It is also necessary to account for the boundary values, which remain constant over time. 
Migration of Jacobi method on the AN is possible by using \textit{conv2D} operation to compute the updated temperature for each point. This computation is combined with a precomputed mask (generated outside the model) to ensure the correct values of the grid are maintained at each iteration. Specifically, a kernel was created to calculate the new value of each point based on its immediate neighbors, thereby recreating a Stencil-5 operation. The mask acts as a kernel in image processing convolution, so a $3 \times 3$ matrix with a value of $0.25$ at the relevant positions can be used with the \textit{torch.tensor} definition (see Listing~\ref{lst:jacobi}). After applying the convolution, the result is multiplied by the precomputed mask to preserve the boundary values. 

\begin{lstlisting}[style=pythonstyle, caption={Jacobi Implementation}, label={lst:jacobi}]
class JacobiMachine(nn.Module):
    def __init__(self, nt=1000, datatype=torch.float32):
        super(JacobiMachine, self).__init__()
        self.datatype=datatype
        self.nt = torch.tensor(nt, dtype=self.datatype)

    def forward(self, X, X_prev, Mask):
      x = X
      x_prev = X_prev
      mask = Mask
      kernel = torch.tensor([[0.0, 0.25, 0.0],
                            [0.25, 0.0, 0.25],
                            [0.0, 0.25, 0.0]], dtype=self.datatype).view(1, 1, 3, 3)
                            
      i = torch.tensor(0, dtype=self.datatype)
      
      while torch.ne(i, self.nt):
          x_prev = x.clone()
          x_next = F.conv2d(x_prev, kernel, padding=1)
          x = x_next * mask
          diff = torch.max(torch.abs(x - x_prev))         
          i = torch.add(i, 1)
      return x
\end{lstlisting}



\subsection{Multigrid}
Among the selected algorithms, the Multigrid implementation was the most complex. The model was designed following a V-Cycle structure, which can be divided into three main phases: descent, coarse grid, and ascent. In addition to these phases, it is necessary to preserve boundary values, similar to the Jacobi method. 


\subsubsection{V-cycle descent}
\label{ssect:descent}
The descent phase requires two primary operations: smoothing and residual computation. Smoothing is performed using one iteration of the Jacobi method described previously, and the residual error of the smoothed solution must be computed. The implementation is described in the Listing~\ref{lst:residuoMult}. 



\begin{lstlisting}[style=pythonstyle, caption={V-cycle descent}, label={lst:residuoMult}]
# ...
# V-cycle descent
for level in range(1, num_levels):
    # ... (mask operations)
    masked_output = F.conv2d(masked_input, kernel, padding=1)
    unmasked_output = F.conv2d(unmasked_input, kernel, padding=1)
    residual = masked_output + unmasked_output
# ...
\end{lstlisting}

The final step of the descent phase is the restriction of the grid. On the ANE, this restriction is implemented using the \textit{Average Pool 2D} layer commonly employed in AI models. The implementation of this operation is illustrated in Listing~\ref{lst:restMult}.

\begin{lstlisting}[style=pythonstyle, caption={Restriction}, label={lst:restMult}]
def restriction(self, residual):
    return nn.AvgPool2d(kernel_size=2)(residual)
\end{lstlisting}

This restriction is applied to the residual computed before, as we can see in Listing \ref{lst:restMultApl}.

\begin{lstlisting}[style=pythonstyle, caption={Restriction usage}, label={lst:restMultApl}]
# ... (Inside the descent loop)
coarse_residual = self.restriction(residual)
grids.append(coarse_residual)
\end{lstlisting}

\subsubsection{Resolution of the Coarsest Grid}
Once the grid has been restricted through all desired levels, the coarsest grid must be solved iteratively. To accomplish this, we apply the Jacobi method. The Listing~\ref{lst:mallaGruesa} illustrates the implementation and computation performed on the coarsest grid.

\begin{lstlisting}[style=pythonstyle, caption={Resolution of coarsest grid}, label={lst:mallaGruesa}]
coarse_solution = self.jacobi(coarse_solution, Mask9)
grids[-1] = coarse_solution
\end{lstlisting}

\subsubsection{V-cycle Ascent}
In the ascent phase, it is necessary to perform a prolongation to restore the grid to its original size. This prolongation also requires a correction to account for the errors introduced at each level due to the coarser grid approximations. The prolongation is implemented using bilinear interpolation through the \textit{interpolate} function in PyTorch. This function computes the value of each new point based on its four nearest neighbors, effectively increasing the grid size at each step. The implementation of this procedure is shown in Listing~\ref{lst:interPol}.

\begin{lstlisting}[style=pythonstyle, caption={Interpolation code}, label={lst:interPol}]
def interpolate(self, f, target_size):
    return F.interpolate(f, size=target_size, mode='bilinear', align_corners=False)
\end{lstlisting}

After the prolongation is applied, a correction must be performed. The implementation of this correction is shown in the Listing~\ref{lst:correc}.

\begin{lstlisting}[style=pythonstyle, caption={Correction applied to the prolonged grid}, label={lst:correc}]
# Upward phase
for level in range(num_levels - 2, -1, -1):
    target_size = grids[level].shape[-2:]
    
    # Interpolate to finer grid
    fine_solution = self.interpolate(grids[level + 1], target_size=target_size)
    
    # Add correction to finer grid
    fine_solution = torch.add(fine_solution, grids[level])
    
    # ... (post-smoothing)
\end{lstlisting}

Finally, a smoothing step is applied to refine the solution. This smoothing procedure is the Jacobi implementation.

\section{Results}
\label{results}
\subsection{Experimental Environment}

The platforms based on Apple's SoC chosen to evaluate both performance and power efficiency are the M1 and M4 Pro Mac-Mini devices, whose main features are summarized in the table~\ref{tab:architectures}.

From the software side, all machine learning models were developed and implemented using the PyTorch framework (v2.7.1). For deployment on Core ML, models were converted using CoreMLTools (v8.3).

\subsection{Experimental Results}

This subsection presents a comparative analysis of performance and energy consumption across a diverse set of computational tasks. Specifically, we report results for two distinct AI models—YOLOv3 and YOLOv11-alongside several general-purpose applications: the standard matrix multiplication, the Jacobi iterative method, and a more complex multigrid solver for the heat equation.

Table~\ref{table:all_sizes} summarizes the problem sizes used for all experiments.

\begin{table}[htbp]
\centering
\caption{Sizes tested on different benchmarks}
\begin{tabular}{llcc}
\toprule
\textbf{Benchmark} & \textbf{Size} & \textbf{M1} & \textbf{M4 Pro} \\
\midrule
\multirow{9}{*}{GEMM} 
 & 256$\times$256 & X & X \\ 
 & 512$\times$512 & X & X \\
 & 1024$\times$1024 & X & X \\
 & 2048$\times$2048 & X & X \\
 & 4096$\times$4096 & X & X \\
 & 8192$\times$8192 & X & X \\
 & 12288$\times$12288 & - & X \\
 & 14336$\times$14336 & - & X \\
 & 16384$\times$16384 & - & X \\
\midrule
\multirow{13}{*}{Jacobi}
 & 256$\times$256 & X & X \\
 & 512$\times$512 & X & X \\
 & 1024$\times$1024 & X & X \\
 & 2048$\times$2048 & X & X \\
 & 4096$\times$4096 & X & X \\
 & 6144$\times$6144 & X & X \\
 & 8192$\times$8192 & X & X \\
 & 10240$\times$10240 & X & X \\
 & 12288$\times$12288 & X & X \\
 & 16384$\times$16384 & - & X \\
 & 32768$\times$32768 & - & X \\
\midrule
\multirow{9}{*}{Multigrid}
 & 512$\times$512 & X & X \\
 & 1024$\times$1024 & X & X \\
 & 2048$\times$2048 & X & X \\
 & 4096$\times$4096 & X & X \\
 & 8192$\times$8192 & X & X \\
 & 10240$\times$10240 & X & X \\
 & 12288$\times$12288 & X & X \\
 & 16384$\times$16384 & X & X \\
 & 32768$\times$32768 & - & X \\
\bottomrule
\end{tabular}
\label{table:all_sizes}
\end{table}

\subsubsection{YOLOv3}
Table~\ref{table:tiempoYolo3} presents the inference time for a single image processed by the YOLOv3 model. Due to the model's small scale, the execution times were exceedingly short (less than 15 ms per inference), which inherently limited the statistical conclusiveness of performance comparisons.

\begin{table}[htbp]
    \centering
    \caption{Execution time for YOLOv3}
    \begin{tabular}{lccc}
        \toprule
        \textbf{Device} & \textbf{CPU (ms)} & \textbf{GPU (ms)} & \textbf{ANE (ms)} \\
        \midrule
        M1      & 14.6 & 8.8 & 5.4 \\
        M4 Pro  & 6.9  & 6.3 & 3.2 \\
        \bottomrule
    \end{tabular}
    \label{table:tiempoYolo3}
\end{table}

Despite the limitations of the model's execution time, some interesting observations can be drawn regarding energy consumption figures (see Table~\ref{table:consumoYolo3}). On the M1 chip, the GPU mode registers the lowest energy consumption, while the ANE mode is the most energy efficient on the M4 Pro chip.

\begin{table}[htbp]
    \centering
    \caption{Mean energy consumption on YOLOv3 (100 executions)}
    \begin{tabularx}{\linewidth}{lXXX}
        \toprule
        \textbf{Device} & \textbf{CPU (mW)} & \textbf{GPU (mW)} & \textbf{ANE (mW)} \\
        \midrule
        M1      & 6237.57 & 4013.45 & 5148.86 \\
        M4 Pro  & 7213.29 & 5523.06 & 5147.58 \\
        \bottomrule
    \end{tabularx}
    \label{table:consumoYolo3}
\end{table}

\subsection{YOLOv11}
In order to obtain more statistically conclusive results, we replicated the experimentation using a newer and larger model, such as YOLOv11. We evaluated the inference time (see Table~\ref{table:tiempo_yolo11_unificada}) for all available YOLOv11 variants: YOLOv11n (2.8M parameters), YOLOv11s (11.7M parameters), YOLOv11m (25.9M parameters), YOLOv11l (43.6M parameters), and YOLOv11x (68.9M parameters).

\begin{table}[htbp]
\centering
\caption{Execution time for YOLOv11}
\begin{tabularx}{\linewidth}{l l l XXX}
\toprule
\textbf{Device} & \textbf{Model} & \textbf{Data Type} & \textbf{CPU (s)} & \textbf{GPU (s)} & \textbf{ANE (s)} \\
\midrule
\multirow{10}{*}{M1} 
 & \multirow{2}{*}{YOLOv11x} & FP16 & 0.6125 & 0.1210 & 0.0327 \\
 & & FP32 & 0.8153 & 0.1383 & - \\
\cmidrule(lr){2-6}
 & \multirow{2}{*}{YOLOv11l} & FP16 & 0.3157 & 0.0581 & 0.0163 \\
 & & FP32 & 0.3777 & 0.0635 & - \\
\cmidrule(lr){2-6}
 & \multirow{2}{*}{YOLOv11m} & FP16 & 0.2760 & 0.0473 & 0.0139 \\
 & & FP32 & 0.3351 & 0.0518 & - \\
\cmidrule(lr){2-6}
 & \multirow{2}{*}{YOLOv11s} & FP16 & 0.0999 & 0.0181 & 0.0055 \\
 & & FP32 & 0.1041 & 0.0188 & - \\
\cmidrule(lr){2-6}
 & \multirow{2}{*}{YOLOv11n} & FP16 & 0.0438 & 0.0082 & 0.0033 \\
 & & FP32 & 0.0412 & 0.0085 & - \\
\midrule
\multirow{10}{*}{M4 Pro} 
 & \multirow{2}{*}{YOLOv11x} & FP16 & 0.0697 & 0.0390 & 0.0222 \\
 & & FP32 & 0.1800 & 0.0416 & - \\
\cmidrule(lr){2-6}
 & \multirow{2}{*}{YOLOv11l} & FP16 & 0.0463 & 0.0193 & 0.0109 \\
 & & FP32 & 0.0861 & 0.0205 & - \\
\cmidrule(lr){2-6}
 & \multirow{2}{*}{YOLOv11m} & FP16 & 0.0348 & 0.0155 & 0.0092 \\
 & & FP32 & 0.0709 & 0.0161 & - \\
\cmidrule(lr){2-6}
 & \multirow{2}{*}{YOLOv11s} & FP16 & 0.0190 & 0.0063 & 0.0031 \\
 & & FP32 & 0.0282 & 0.0066 & - \\
\cmidrule(lr){2-6}
 & \multirow{2}{*}{YOLOv11n} & FP16 & 0.0095 & 0.0032 & 0.0016 \\
 & & FP32 & 0.0132 & 0.0032 & - \\
\bottomrule
\end{tabularx}
\label{table:tiempo_yolo11_unificada}
\end{table}

The obtained results confirm the expected performance hierarchy: the ANE, designed specifically for AI workloads, consistently yields the fastest inference times, followed by the GPU. However, a significant generational shift is observed in the acceleration margin between the M1 and M4 Pro chips. Focusing on ANE vs CPU, the M1 chip demonstrated a substantial speedup of the ANE over the CPU, ranging from $13\times$ to $20\times$. In contrast, the M4 Pro shows a much narrower gain, with acceleration factors ranging only from $3\times$ to $6\times$. From the GPU vs. CPU analysys, the narrowing of the performance gap is even more pronounced: M1's ANE provided a speedup from $3\times$ to $5\times$ which is reduced to $1.7\times$ to $2.0\times$ on the M4 Pro. These results point to a substantially improved baseline performance on the M4 Pro's CPU and GPU, suggesting that the ANE may have reached a performance ceiling for this specific inference task.

Regarding energy consumption, the results for the YOLOv11 models are summarized in Table~\ref{table:consumo_yolo11_unificada}.

\begin{table}[htbp]
\centering
\caption{Mean energy consumption on YOLOv11 (100 executions)}
\begin{tabularx}{\linewidth}{l l l XXX}
\toprule
\textbf{Device} & \textbf{Model} & \textbf{Datatype} & \textbf{CPU (mW)} & \textbf{GPU (mW)} & \textbf{ANE (mW)} \\
\midrule
\multirow{10}{*}{M1} 
 & \multirow{2}{*}{YOLOv11x} & FP16 & 5682.03 & 8353.86 & 5884.40 \\
 & & FP32 & 5647.42 & 8060.65 & - \\
\cmidrule(lr){2-6}
 & \multirow{2}{*}{YOLOv11l} & FP16 & 5693.44 & 7399.11 & 5706.76 \\
 & & FP32 & 5913.78 & 7600.92 & - \\
\cmidrule(lr){2-6}
 & \multirow{2}{*}{YOLOv11m} & FP16 & 5694.54 & 7379.86 & 5735.29 \\
 & & FP32 & 5660.16 & 7692.03 & - \\
\cmidrule(lr){2-6}
 & \multirow{2}{*}{YOLOv11s} & FP16 & 5949.13 & 6135.68 & 5276.77 \\
 & & FP32 & 6111.55 & 6703.50 & - \\
\cmidrule(lr){2-6}
 & \multirow{2}{*}{YOLOv11n} & FP16 & 6078.40 & 5380.35 & 5669.67 \\
 & & FP32 & 5982.78 & 5924.13 & - \\
\midrule
\multirow{10}{*}{M4 Pro} 
 & \multirow{2}{*}{YOLOv11x} & FP16 & 7657.90 & 11095.23 & 8452.22 \\
 & & FP32 & 7332.38 & 12715.25 & - \\
\cmidrule(lr){2-6}
 & \multirow{2}{*}{YOLOv11l} & FP16 & 7684.67 & 10871.15 & 8769.38 \\
 & & FP32 & 7879.98 & 16868.31 & - \\
\cmidrule(lr){2-6}
 & \multirow{2}{*}{YOLOv11m} & FP16 & 7846.11 & 12405.33 & 9644.23 \\
 & & FP32 & 7892.85 & 13474.40 & - \\
\cmidrule(lr){2-6}
 & \multirow{2}{*}{YOLOv11s} & FP16 & 7740.40 & 10444.30 & 6993.56 \\
 & & FP32 & 8294.97 & 14246.92 & - \\
\cmidrule(lr){2-6}
 & \multirow{2}{*}{YOLOv11n} & FP16 & 8414.38 & 10578.93 & 10415.67 \\
 & & FP32 & 9313.65 & 10339.00 & - \\
\bottomrule
\end{tabularx}
\label{table:consumo_yolo11_unificada}
\end{table}

Energy consumption is markedly higher on the M4 Pro, confirming that the performance gains in generational Apple Silicon come at a power price. The GPU consistently registers the highest power usage on both the M1 and M4 Pro. Taking into account the energy perspective, the ANE is still the more efficient device.

The markedly higher energy consumption on the M4 Pro confirms that the generational performance gains come at increased power budget. The GPU is the most power-intensive unit. Despite these variations in raw power metrics, when factoring in the fastest inference speed, the ANE remains the most energy-efficient option overall.

\subsection{GEMM}
To effectively evaluate the performance of the GEMM-based benchmark, we must first consider certain aspects of the Apple Silicon architecture. This architecture incorporates a dedicated unit called the AMX (Apple Matrix Coprocessor) available within each multi-core cluster. This coprocessor offloads intensive matrix operations, thus freeing the main CPU cores. Programming the AMX is done through high-level Apple libraries, primarily the Accelerate framework or Core ML. Therefore, we would lke to evaluate the behaviour of GEMM by means of both programming model on CPU devices. To do this, we implemented a baseline C++ code using the Accelerate framework to evaluate performance differences between CPU-only and AMX-enabled executions. Accelerate framework can be easily used for GEMM with his own implementation\footnote{\url{https://developer.apple.com/documentation/accelerate/cblas_sgemm(_:_:_:_:_:_:_:_:_:_:_:_:_:_:)}}, we just need to call the method call \texttt{cblas\_sgemm} like is shown in the Listing \ref{lst:Acc}.

\begin{lstlisting}[style=pythonstyle, caption={Using AMX for Matrix Multiplication via Accelerate}, label={lst:Acc}]
#include <vector>
#include <Accelerate/Accelerate.h>

int main() {
    int M, N, K; //Dimensions
    int lda=K, ldb=N, ldc=N;

    // Init A, B, C
    std::vector<float> A(M*K, 2.0f); 
    std::vector<float> B(K*N, 1.0f); 
    std::vector<float> C(M*N, 0.0f);

    // Exec C = 1.0 * A * B + 0.0 * C
    cblas_sgemm(CblasRowMajor, CblasNoTrans, 
        CblasNoTrans, M, N, K, 
                1.0f, A.data(), lda,
                B.data(), ldb, 
                0.0f, C.data(), ldc); 
    return 0;
}
\end{lstlisting}

Table \ref{tab:rendimiento_comparativa} analyzes the GEMM execution times, comparing a baseline C++ implementation and the Accelerate library with the performance achieved by Core ML targeting the CPU device in both M1 and M4 Pro systems.

\begin{table}[htbp]
    \centering
    \caption{GEMM performance on CPU M1 and M4 Pro (GFLOPS)}
    \begin{tabularx}{\linewidth}{c c X X X}
        \toprule
        \textbf{Matrix size (k)} & \textbf{Device} & \makecell{\textbf{Accelerate}\\(1 thread)} & \makecell{\textbf{Accelerate}\\ (2 threads)} & \makecell{\textbf{CoreML}} \\
        \midrule
        \multirow{2}{*}{256}   & M1     & 328.29   & -       & 369.39   \\
                               & M4 Pro & 564.33   & 462.87  & 611.90   \\
        \midrule
        \multirow{2}{*}{512}   & M1     & 745.05   & -       & 715.77   \\
                               & M4 Pro & 1120.62  & 819.23  & 1376.41  \\
        \midrule
        \multirow{2}{*}{1024}  & M1     & 980.57   & -       & 606.63   \\
                               & M4 Pro & 1257.37  & 1378.69 & 1356.71  \\
        \midrule
        \multirow{2}{*}{2048}  & M1     & 843.57   & -       & 811.93   \\
                               & M4 Pro & 1600.02  & 1599.76 & 1599.15  \\
        \midrule
        \multirow{2}{*}{4096}  & M1     & 923.31   & -       & 1032.54  \\
                               & M4 Pro & 1584.81  & 2546.39 & 1528.89  \\
        \midrule
        \multirow{2}{*}{8192}  & M1     & 870.99   & -       & 835.55   \\
                               & M4 Pro & 1542.35  & 3016.10 & 1609.28  \\
        \midrule
        \multirow{2}{*}{12288} & M1     & -        & -       & -        \\
                               & M4 Pro & 1536.88  & 3041.88 & 1648.19  \\
        \midrule
        \multirow{2}{*}{14336} & M1     & -        & -       & -        \\
                               & M4 Pro & 1534.89  & 3039.61 & 1589.69  \\
        \midrule
        \multirow{2}{*}{16384} & M1     & -        & -       & -        \\
                               & M4 Pro & 1529.61  & 3069.36 & 1591.63  \\
        \bottomrule
    \end{tabularx}
    \label{tab:rendimiento_comparativa}
\end{table}

The first observation from the results (see Table \ref{tab:rendimiento_comparativa}) is rooted in the architecture: the M4 Pro incorporates two AMX accelerators due to its two P-core clusters, as reported by some authors \cite{Jonathan_thesis2025}. In contrast, the M1 platform utilizes only one AMX unit. A second key finding is that the GEMM operation implemented via Core ML Tools defaults to the AMX accelerator when the CPU device is selected using \textit{ct.ComputeUnit.CPU\_ONLY}. However, Core ML does not allow the exploitation of both accelerators available in the M4 Pro.

Focusing on GEMM performance on the M1 with CoreML, as shown in Figure~\ref{fig:m1matmulPerformance}, the ANE consistently achieves the highest performance for matrix sizes greater than or equal to 1024, but its usage is limited to $4096^3$ matrix sizes. For smaller matrices, the CPU performs better due to AMX utilization. 
\begin{figure}[htbp]
    \centering
    \caption{GEMM performance on Mac M1}
    \begin{tikzpicture}
        \begin{axis}[
            width=\linewidth,       
            height=0.55\linewidth,  
            xlabel={$k$ Dimension},
            ylabel={GFLOPS},
            legend style={
                at={(0.5,-0.33)},
                anchor=north,
                legend columns=2,
                font=\small
            },
            grid=both,
            xmin=0, xmax=8200,
            ymin=0, ymax=2800,
            tick label style={font=\small},
            label style={font=\small},
            title style={font=\small\bfseries},
            legend cell align={left}
        ]

        \addplot+[mark=x, color=cyan] coordinates {
            (0,0) (256,246.91) (512,513.41) (1024,497.36) (2048,723.24) (4096,953.77) (8192,833.63)
        };
        \addplot+[mark=*, color=RoyalPurple] coordinates {
            (0,0) (256,369.39) (512,715.77) (1024,606.63) (2048,811.93) (4096,1032.54) (8192,835.55) 
        };
        \addplot+[mark=square*, color=ForestGreen] coordinates {
            (0,0) (256,91.93) (512,302.58) (1024,508.65) (2048,1044.89) (4096,1314.78) (8192,1529.33)
        };
        \addplot+[mark=square, color=BurntOrange] coordinates {
            (0,0) (256,96.93) (512,373.93) (1024,587.06) (2048,1152.31) (4096,1303.55) (8192,1456.07)
        };
        \addplot+[mark=o, color=BrickRed] coordinates {
            (0,0) (256,94.26) (512,406.02) (1024,911.84) (2048,1659.74) (4096,2567.23)
        };

        \legend{M1 AMX FP16, M1 AMX FP32, M1 GPU FP16, M1 GPU FP32, M1 ANE FP16}
        \end{axis}
    \end{tikzpicture}
    \label{fig:m1matmulPerformance}
\end{figure}

The performance results achieved on the M4 Pro system  are shown in the Figure~\ref{fig:m4matmulPerformance}. The results for the M4 Pro differ significantly from those observed on the M1. In this experiment, the ANE does not achieve the highest performance, although it performs successfully for smaller matrix sizes. The GPU emerges as the leading device, demonstrating substantial improvement compared to the M1. Additionally, the performance of the ANE decreases for matrices sizes from $4096^3$ and larger, likely because the limitation on the ANE internal memory buffer.

\begin{figure}[htbp]
    \centering
    \caption{GEMM performance on Mac M4 Pro}
    \begin{tikzpicture}
        \begin{axis}[
            width=\linewidth,        
            height=0.55\linewidth,   
            xlabel={Dimensión $k$},
            ylabel={GFLOPS},
            legend style={
                at={(0.5,-0.33)},
                anchor=north,
                legend columns=2,
                font=\small
            },
            grid=both,
            xmin=0, xmax=16500,
            ymin=0, ymax=5000,
            tick label style={font=\small},
            label style={font=\small},
            title style={font=\small\bfseries},
            legend cell align={left},
            scaled ticks=false,
            xticklabel style={/pgf/number format/fixed, /pgf/number format/1000 sep={\,}}
        ]

        \addplot+[mark=x, color=cyan] coordinates {
            (0,0) (256,332.71) (512,813.51) (1024,1245.64) (2048,1953.15) (4096,1411.71) (8192,1133.46) (12288,1166.33) (14336,1222.94) (16384,1186.01)
        };
        \addplot+[mark=*, color=RoyalPurple] coordinates {
            (0,0) (256,611.90) (512,1376.41) (1024,1356.71) (2048,1599.15) (4096,1528.89) (8192,1609.28) (12288,1648.19) (14336,1589.69) (16384,1591.63)
        };
        \addplot+[mark=square*, color=ForestGreen] coordinates {
            (0,0) (1024,1302.37) (2048,2531.36) (4096,3161.34) (8192,4058.29) (12288,4381.79) (14336,4425.03) (16384,3161.34)
        };
        \addplot+[mark=square, color=BurntOrange] coordinates {
            (0,0) (1024,2170.93) (2048,3670.79) (4096,4078.42) (8192,4491.05) (12288,4745.69) (14336,4696.08) (16384,4748.03)
        };
        \addplot+[mark=o, color=BrickRed] coordinates {
            (0,0) (1024,1398.20) (2048,3263.77) (4096,3806.63) (8192,3114.22) (12288,1359.95) (14336,1213.40)
        };

        \legend{M4 Pro AMX FP16, M4 Pro AMX FP32, M4 Pro GPU FP16, M4 Pro GPU FP32, M4 Pro ANE FP16}
        \end{axis}
    \end{tikzpicture}
    \label{fig:m4matmulPerformance}
\end{figure}

\begin{figure}[htbp]
    \centering
    \caption{GEMM energy efficiency on Mac M1}
    \begin{tikzpicture}
        \begin{axis}[
            width=\linewidth,        
            height=0.55\linewidth,   
            xlabel={Dimensión $k$},
            ylabel={GFLOPS/W},
            legend style={
                at={(0.5,-0.33)},
                anchor=north,
                legend columns=2,
                font=\small
            },
            grid=both,
            xmin=0, xmax=8200,
            ymin=0, ymax=300,
            tick label style={font=\small},
            label style={font=\small},
            title style={font=\small\bfseries},
            legend cell align={left}
        ]

        \addplot+[mark=x, color=cyan] coordinates {
            (0,0) (256,48.6151) (512,97.6213) (1024,95.1469) (2048,126.2102) (4096,150.7282) (8192,153.6237)
        };
        \addplot+[mark=*, color=RoyalPurple] coordinates {
            (0,0) (256,73.7324) (512,133.5267) (1024,116.6212) (2048,141.3540) (4096,160.0882) (8192,154.3243)
        };
        \addplot+[mark=square*, color=ForestGreen] coordinates {
            (0,0) (256,20.9097) (512,90.5035) (1024,268.6786) (2048,248.0100) (4096,221.6440) (8192,210.2564)
        };
        \addplot+[mark=square, color=BurntOrange] coordinates {
            (0,0) (256,21.8114) (512,92.2790) (1024,268.5783) (2048,232.6582) (4096,196.5563) (8192,181.3376)
        };
        \addplot+[mark=o, color=BrickRed] coordinates {
            (0,0) (256,20.7364) (512,84.1786) (1024,157.4170) (2048,196.3193) (4096,252.8610)
        };

        \legend{M1 CPU FP16, M1 CPU FP32, M1 GPU FP16, M1 GPU FP32, M1 ANE FP16}
        \end{axis}
    \end{tikzpicture}
    \label{fig:m1matmulEficiencia}
\end{figure}
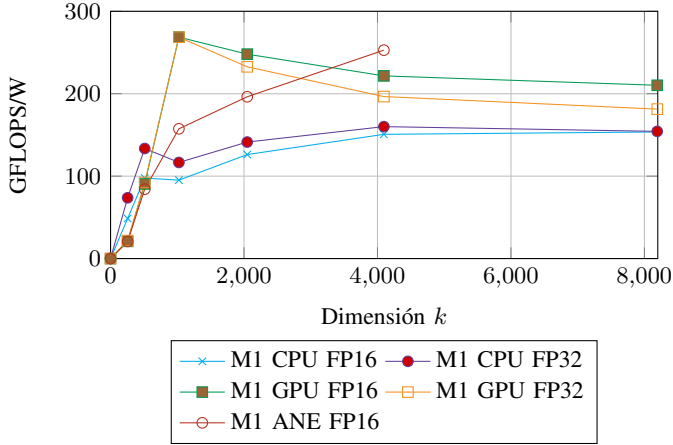

Regarding energy efficiency, the results for the M1 are presented in Figure~\ref{fig:m1matmulEficiencia}. For the largest matrix size tested (4096), the ANE demonstrates the highest energy efficiency. For smaller sizes, the GPU exhibits the best efficiency, particularly when using FP16 data. The CPU also shows notable efficiency benefits, largely attributable to the utilization of the AMX.

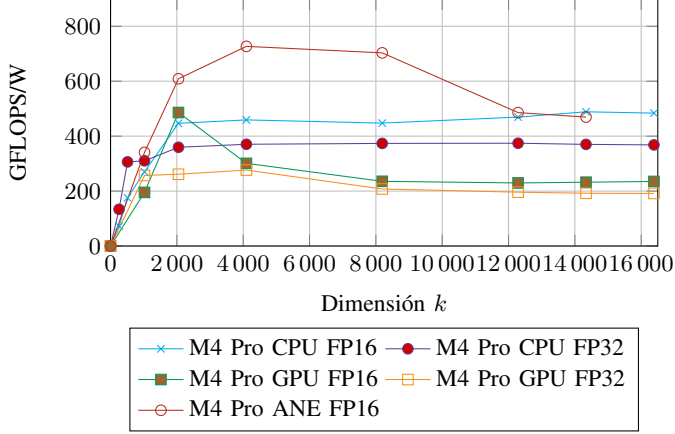
\begin{figure}[htbp]
    \centering
    \caption{GEMM energy efficiency on Mac M4 Pro}
    \begin{tikzpicture}
        \begin{axis}[
            width=\linewidth,        
            height=0.55\linewidth,   
            xlabel={Dimensión $k$},
            ylabel={GFLOPS/W},
            legend style={
                at={(0.5,-0.33)},
                anchor=north,
                legend columns=2,
                font=\small
            },
            grid=both,
            xmin=0, xmax=16500,
            ymin=0, ymax=900,
            tick label style={font=\small},
            label style={font=\small},
            title style={font=\small\bfseries},
            legend cell align={left},
            scaled ticks=false,
            xticklabel style={/pgf/number format/fixed, /pgf/number format/1000 sep={\,}}
        ]

        \addplot+[mark=x, color=cyan] coordinates {
            (0,0) (256,72.3998) (512,175.7245) (1024,270.1633) (2048,446.6255) (4096,459.0890) (8192,447.4617) (12288,469.3627) (14336,488.5810) (16384,483.8766)
        };
        \addplot+[mark=*, color=RoyalPurple] coordinates {
            (0,0) (256,133.8013) (512,306.2773) (1024,310.1613) (2048,359.4139) (4096,370.3989) (8192,373.5370) (12288,374.2339) (14336,370.0007) (16384,368.1108)
        };
        \addplot+[mark=square*, color=ForestGreen] coordinates {
            (0,0) (1024,195.2704) (2048,485.9670) (4096,300.9515) (8192,235.5465) (12288,229.5624) (14336,232.2746) (16384,235.0232)
        };
        \addplot+[mark=square, color=BurntOrange] coordinates {
            (0,0) (1024,256.8054) (2048,261.2862) (4096,276.4676) (8192,207.5804) (12288,196.0010) (14336,192.0737) (16384,191.4285)
        };
        \addplot+[mark=o, color=BrickRed] coordinates {
            (0,0) (1024,340.7865) (2048,608.8763) (4096,726.7927) (8192,703.1185) (12288,485.6721) (14336,469.4253)
        };

        \legend{M4 Pro CPU FP16, M4 Pro CPU FP32, M4 Pro GPU FP16, M4 Pro GPU FP32, M4 Pro ANE FP16}
        \end{axis}
    \end{tikzpicture}
    \label{fig:m4matmulEficiencia}
\end{figure}

The energy efficiency results for the M4 Pro are shown in Figure~\ref{fig:m4matmulEficiencia}. In most of configurations, the ANE demonstrates the highest efficiency. This figure also highlights that the substantial performance gains of the GPU come at the cost of significantly higher energy consumption. Additionally, the CPU maintains relatively good efficiency, which can be attributed to the utilization of the AMX.

\subsection{Jacobi}

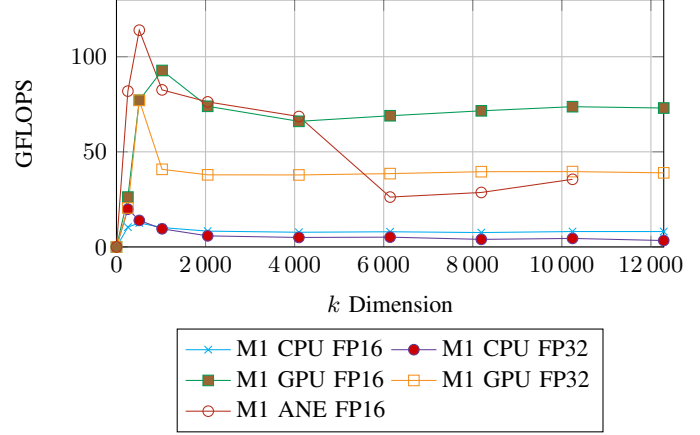
\begin{figure}[htbp]
    \centering
    \caption{Jacobi Performance on M1}
    \begin{tikzpicture}
        \begin{axis}[
            width=\linewidth,        
            height=0.55\linewidth,   
            xlabel={$k$ Dimension},
            ylabel={GFLOPS},
            legend style={
                at={(0.5,-0.33)},
                anchor=north,
                legend columns=2,
                font=\small
            },
            grid=both,
            xmin=0, xmax=12288,
            ymin=0, ymax=130,
            tick label style={font=\small},
            label style={font=\small},
            title style={font=\small\bfseries},
            legend cell align={left},
            scaled ticks=false,
            xticklabel style={/pgf/number format/fixed, /pgf/number format/1000 sep={\,}}
        ]

        \addplot+[mark=x, color=cyan] coordinates {
            (0,0) (256,10.4) (512,12.61) (1024,10.18) (2048,8.31) (4096,7.71) (6144,7.97) (8192,7.56) (10240,8.09) (12288,8.09)
        };
        \addplot+[mark=*, color=RoyalPurple] coordinates {
            (0,0) (256,19.86) (512,13.94) (1024,9.52) (2048,5.81) (4096,4.98) (6144,5.14) (8192,3.97) (10240,4.47) (12288,3.36)
        };
        \addplot+[mark=square*, color=ForestGreen] coordinates {
            (0,0) (256,26.21) (512,77.10) (1024,92.77) (2048,73.94) (4096,66.03) (6144,68.97) (8192,71.57) (10240,73.73) (12288,73.03)
        };
        \addplot+[mark=square, color=BurntOrange] coordinates {
            (0,0) (256,20.48) (512,77.10) (1024,40.80) (2048,37.91) (4096,37.85) (6144,38.54) (8192,39.55) (10240,39.61) (12288,38.95)
        };
        \addplot+[mark=o, color=BrickRed] coordinates {
            (0,0) (256,81.92) (512,113.97) (1024,82.56) (2048,76.26) (4096,68.57) (6144,26.21) (8192,28.63) (10240,35.55)
        };

        \legend{M1 CPU FP16, M1 CPU FP32, M1 GPU FP16, M1 GPU FP32, M1 ANE FP16}
        \end{axis}
    \end{tikzpicture}
    \label{fig:m1jacobiPerformance}
\end{figure}

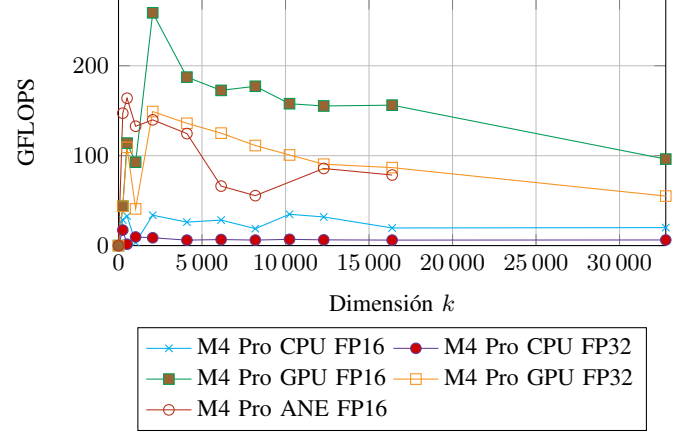
\begin{figure}[htbp]
    \centering
    \caption{Jacobi Performance on M4 Pro}
    \begin{tikzpicture}
        \begin{axis}[
            width=\linewidth,        
            height=0.55\linewidth,   
            xlabel={Dimensión $k$},
            ylabel={GFLOPS},
            legend style={
                at={(0.5,-0.33)},
                anchor=north,
                legend columns=2,
                font=\small
            },
            grid=both,
            xmin=0, xmax=32768,
            ymin=0, ymax=275,
            tick label style={font=\small},
            label style={font=\small},
            title style={font=\small\bfseries},
            legend cell align={left},
            scaled ticks=false,
            xticklabel style={/pgf/number format/fixed, /pgf/number format/1000 sep={\,}}
        ]

        \addplot+[mark=x, color=cyan] coordinates {
            (0,0) (256,28.65) (512,32.77) (1024,3.44) (2048,33.79) (4096,26.06) (6144,28.36) (8192,18.83) (10240,34.82) (12288,31.82) (16384,19.59) (32768,19.97) 
        };
        \addplot+[mark=*, color=RoyalPurple] coordinates {
            (0,0) (256,17.22) (512,1.53) (1024,9.52) (2048,8.67) (4096,6.06) (6144,6.71) (8192,6.11) (10240,6.95) (12288,6.37) (16384,6.09) (32768,6.24)
        };
        \addplot+[mark=square*, color=ForestGreen] coordinates {
            (0,0) (256,43.81) (512,113.97) (1024,92.77) (2048,258.89) (4096,187.33) (6144,172.61) (8192,177.18) (10240,157.78) (12288,155.26) (16384,156.17) (32768,96.19)
        };
        \addplot+[mark=square, color=BurntOrange] coordinates {
            (0,0) (256,43.72) (512,109.23) (1024,40.80) (2048,149.08) (4096,135.97) (6144,125.07) (8192,111.27) (10240,100.78) (12288,90.51) (16384,86.70) (32768,55.07)
        };
        \addplot+[mark=o, color=BrickRed] coordinates {
            (0,0) (256,147.23) (512,163.84) (1024,132.72) (2048,139.81) (4096,124.56) (6144,66.14) (8192,55.53) (12288,85.77) (16384,78.47)
        };

        \legend{M4 Pro CPU FP16, M4 Pro CPU FP32, M4 Pro GPU FP16, M4 Pro GPU FP32, M4 Pro ANE FP16}
        \end{axis}
    \end{tikzpicture}
    \label{fig:m4jacobiPerformance}
\end{figure}

Figures \ref{fig:m1jacobiPerformance} and \ref{fig:m4jacobiPerformance} present the performance rates achieved for the Jacobi algorithm on the M1 and M4 Pro, respectively. Due to the more complex computational flow and potential memory access patterns of the Jacobi method, the overall performance rates have significantly dropped compared to the more compute-bound GEMM test. As can be seen, the ANE remains the most powerful device for small grid sizes (less than $1024^2$), while the GPU reports better GFLOPs for larger problem sizes. This performance behavior is consistent across both the M1 and M4 Pro platforms.

Regarding energy efficiency, the Jacobi algorithm exhibits lower ANE efficiency compared to GEMM as it is shown in Figures~\ref{fig:m1jacobiEficiencia} and ~\ref{fig:m4jacobiEficiencia}. On the M4 Pro, ANE is the more efficient device comparing, being more than double efficiency on average (from 20-40 GFLOPs/Watts) in comparison with the GPU counterpart.

\begin{figure}[htbp]
    \centering
    \caption{Jacobi energy efficiency on Mac M1}
    \begin{tikzpicture}
        \begin{axis}[
            width=\linewidth,        
            height=0.55\linewidth,   
            xlabel={Dimensión $k$},
            ylabel={GFLOPS/W},
            legend style={
                at={(0.5,-0.33)},
                anchor=north,
                legend columns=2,
                font=\small
            },
            grid=both,
            xmin=0, xmax=12288,
            ymin=0, ymax=22,
            tick label style={font=\small},
            label style={font=\small},
            legend cell align={left},
            scaled ticks=false,
            xticklabel style={
                /pgf/number format/fixed,
                /pgf/number format/1000 sep={\,}
            }
        ]

        \addplot+[mark=x, color=cyan] coordinates {
            (0,0) (256,1.08) (512,1.85) (1024,1.64) (2048,1.49) (4096,1.44) (6144,1.48) (8192,1.37) (10240,1.47) (12288,1.47)
        };
        \addplot+[mark=*, color=RoyalPurple] coordinates {
            (0,0) (256,2.02) (512,2.19) (1024,1.77) (2048,1.25) (4096,1.00) (6144,1.05) (8192,0.89) (10240,0.98) (12288,0.84)
        };
        \addplot+[mark=square*, color=ForestGreen] coordinates {
            (0,0) (256,2.76) (512,11.19) (1024,14.21) (2048,13.12) (4096,11.37) (6144,11.20) (8192,11.18) (10240,11.12) (12288,10.89)
        };
        \addplot+[mark=square, color=BurntOrange] coordinates {
            (0,0) (256,2.11) (512,10.95) (1024,7.92) (2048,8.66) (4096,8.70) (6144,8.57) (8192,8.39) (10240,8.43) (12288,8.50)
        };
        \addplot+[mark=o, color=BrickRed] coordinates {
            (0,0) (256,14.32) (512,20.83) (1024,15.61) (2048,15.52) (4096,13.65) (6144,5.23) (8192,5.69) (10240,7.15)
        };

        \legend{M1 CPU FP16, M1 CPU FP32, M1 GPU FP16, M1 GPU FP32, M1 ANE FP16}
        \end{axis}
    \end{tikzpicture}
    \label{fig:m1jacobiEficiencia}
\end{figure}
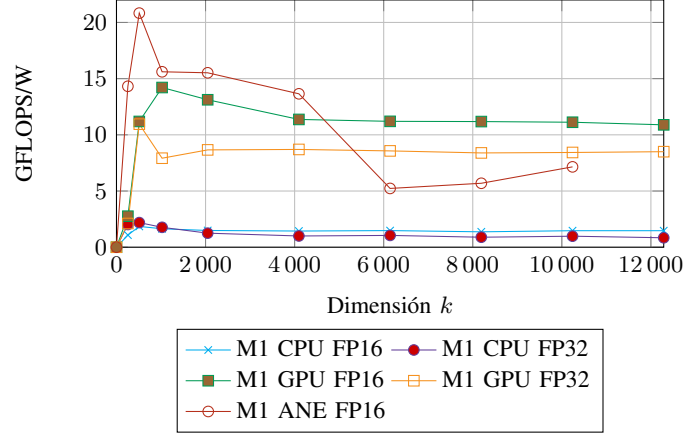


\begin{figure}[htbp]
    \centering
    \caption{Jacobi energy efficiency on M4 Pro}
    \begin{tikzpicture}
        \begin{axis}[
            width=\linewidth,        
            height=0.55\linewidth,   
            xlabel={Dimensión $k$},
            ylabel={GFLOPS/W},
            legend style={
                at={(0.5,-0.33)},
                anchor=north,
                legend columns=2,
                font=\small
            },
            grid=both,
            xmin=0, xmax=32768,
            ymin=0, ymax=45,
            tick label style={font=\small},
            label style={font=\small},
            title style={font=\small\bfseries},
            legend cell align={left},
            scaled ticks=false,
            xticklabel style={
                /pgf/number format/fixed,
                /pgf/number format/1000 sep={\,}
            }
        ]

        \addplot+[mark=x, color=cyan] coordinates {
            (0,0) (256,1.93) (512,2.81) (1024,0.39) (2048,4.11) (4096,4.39) (6144,5.22) (8192,3.31) (10240,5.34) (12288,5.54) (16384,3.41) (32768,3.47)
        };
        \addplot+[mark=*, color=RoyalPurple] coordinates {
            (0,0) (256,1.76) (512,0.15) (1024,1.41) (2048,1.32) (4096,1.40) (6144,1.79) (8192,1.70) (10240,1.67) (12288,1.75) (16384,1.70) (32768,1.75)
        };
        \addplot+[mark=square*, color=ForestGreen] coordinates {
            (0,0) (256,2.59) (512,11.61) (1024,11.29) (2048,29.36) (4096,17.71) (6144,13.68) (8192,14.92) (10240,12.53) (12288,12.30) (16384,11.68) (32768,9.57)
        };
        \addplot+[mark=square, color=BurntOrange] coordinates {
            (0,0) (256,5.17) (512,12.58) (1024,4.56) (2048,14.06) (4096,11.23) (6144,9.87) (8192,8.68) (10240,8.10) (12288,7.37) (16384,7.10) (32768,10.04)
        };
        \addplot+[mark=o, color=BrickRed] coordinates {
            (0,0) (256,19.15) (512,19.28) (1024,25.21) (2048,33.91) (4096,39.61) (6144,33.03) (8192,31.06) (12288,22.26) (16384,37.28)
        };

        \legend{M4 Pro CPU FP16, M4 Pro CPU FP32, M4 Pro GPU FP16, M4 Pro GPU FP32, M4 Pro ANE FP16}
        \end{axis}
    \end{tikzpicture}
    \label{fig:m4jacobiEficiencia}
\end{figure}

\subsection{Multigrid}

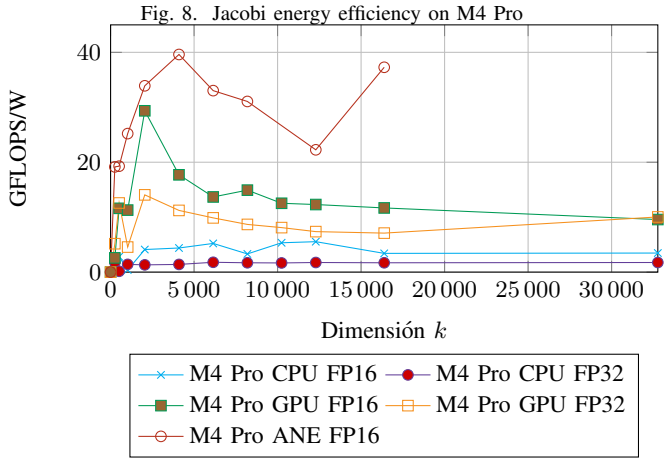
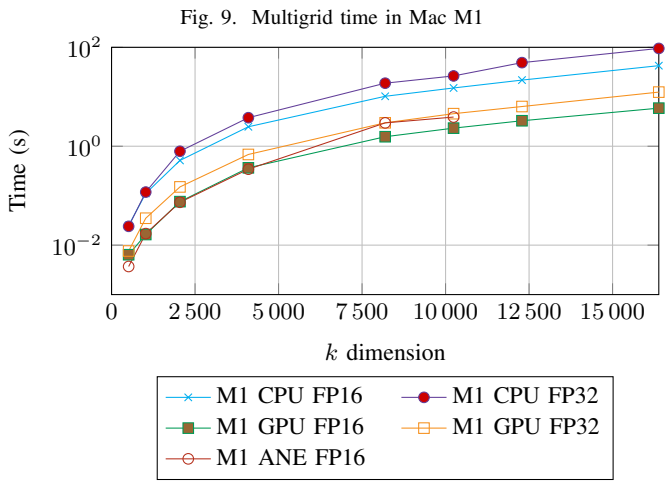
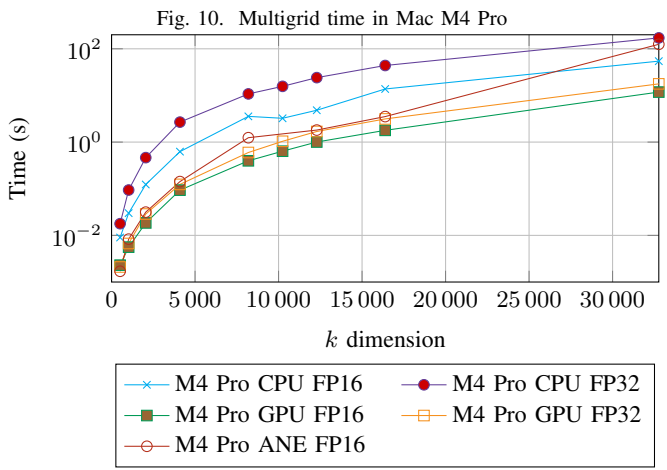
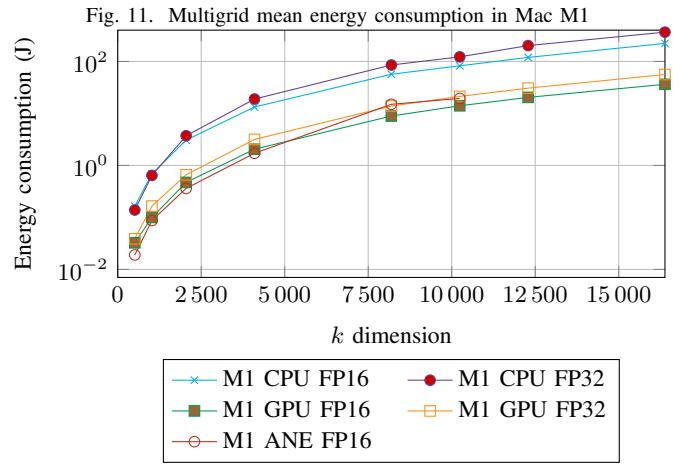
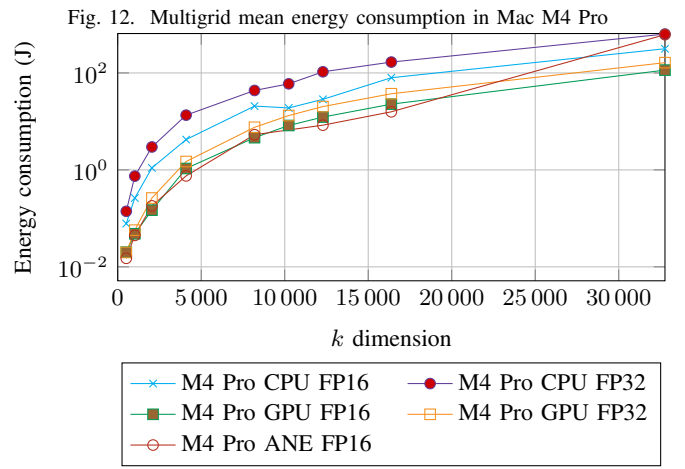
\begin{figure}[htbp]
    \centering
    \caption{Multigrid time in Mac M1}
    \begin{tikzpicture}
        \begin{axis}[
            width=\linewidth,       
            height=0.55\linewidth,  
            xlabel={$k$ dimension},
            ylabel={Time (s)},
            grid=both,
            ymode=log,              
            xmin=0, xmax=16384,
            ymin=1e-3, ymax=100,    
            tick label style={font=\small},
            label style={font=\small},
            title style={font=\small\bfseries},
            legend cell align={left},
            legend style={
                at={(0.5,-0.33)},
                anchor=north,
                legend columns=2,
                font=\small,
                /tikz/every even column/.append style={column sep=0.4cm}
            },
            scaled ticks=false,
            xtick={0, 2500, 5000, 7500, 10000, 12500, 15000},
            xticklabel style={
                /pgf/number format/fixed,
                /pgf/number format/1000 sep={\,}
            }
        ]

        \addplot+[mark=x, color=cyan] coordinates {
            (512,0.025) (1024,0.1112) (2048,0.5161) (4096,2.4760) (8192,10.2296) (10240,15.0757) (12288,21.7556) (16384,42.5806)
        };
        \addplot+[mark=*, color=RoyalPurple] coordinates {
            (512,0.024) (1024,0.1183) (2048,0.7912) (4096,3.7754) (8192,18.8311) (10240,26.3987) (12288,49.0395) (16384,95.0573)
        };
        \addplot+[mark=square*, color=ForestGreen] coordinates {
            (512,0.0064) (1024,0.0164) (2048,0.0758) (4096,0.3658) (8192,1.5570) (10240,2.3223) (12288,3.2754) (16384,5.8897)
        };
        \addplot+[mark=square, color=BurntOrange] coordinates {
            (512,0.0076) (1024,0.0350) (2048,0.1504) (4096,0.6815) (8192,2.9775) (10240,4.5531) (12288,6.3534) (16384,12.4295)
        };
        \addplot+[mark=o, color=BrickRed] coordinates {
            (512,0.0037) (1024,0.0171) (2048,0.0735) (4096,0.3433) (8192,2.9649) (10240,3.8909)
        };

        \legend{M1 CPU FP16, M1 CPU FP32, M1 GPU FP16, M1 GPU FP32, M1 ANE FP16}
        \end{axis}
    \end{tikzpicture}
    \label{fig:m1multigridtime}
\end{figure}

The performance results for the Multigrid algorithm follow the general trend observed with the Jacobi smoother. Figures~\ref{fig:m1multigridtime} and~\ref{fig:m4multigridTime} show the mean execution time of the Multigrid across multiple problem sizes. Taking into account the greater complexity of this method, which involves solving a hierarchy of grid sizes, the GPU is the fastest device (lowest execution time) independent of the problem size. As in previous analyses, the ANE is competitive for small problem sizes, but as the grid size increases, the ANE’s execution time increases significantly, causing its efficiency to drop.

\begin{figure}[htbp]
    \centering
    \caption{Multigrid time in Mac M4 Pro}
    \begin{tikzpicture}
        \begin{axis}[
            width=\linewidth,       
            height=0.55\linewidth,  
            xlabel={$k$ dimension},
            ylabel={Time (s)},
            grid=both,
            ymode=log,              
            xmin=0, xmax=32768,
            ymin=1e-3, ymax=200,    
            tick label style={font=\small},
            label style={font=\small},
            title style={font=\small\bfseries},
            legend cell align={left},
            legend style={
                at={(0.5,-0.33)},
                anchor=north,
                legend columns=2,
                font=\small,
                /tikz/every even column/.append style={column sep=0.4cm}
            },
            scaled ticks=false,
            xtick={0, 5000, 10000, 15000, 20000, 25000, 30000},
            xticklabel style={
                /pgf/number format/fixed,
                /pgf/number format/1000 sep={\,}
            }
        ]

        \addplot+[mark=x, color=cyan] coordinates {
            (512,0.009) (1024,0.030) (2048,0.1233) (4096,0.6246) (8192,3.5600) 
            (10240,3.2388) (12288,4.8176) (16384,13.7581) (32768,54.3670)
        };
        \addplot+[mark=*, color=RoyalPurple] coordinates {
            (512,0.0178) (1024,0.0938) (2048,0.4645) (4096,2.6809) (8192,10.7890) 
            (10240,15.6723) (12288,24.0943) (16384,43.8594) (32768,171.3417)
        };
        \addplot+[mark=square*, color=ForestGreen] coordinates {
            (512,0.0023) (1024,0.0056) (2048,0.0185) (4096,0.0933) (8192,0.3972) 
            (10240,0.6379) (12288,1.0035) (16384,1.7863) (32768,11.8149)
        };
        \addplot+[mark=square, color=BurntOrange] coordinates {
            (512,0.0021) (1024,0.0067) (2048,0.0287) (4096,0.1239) (8192,0.5981) 
            (10240,1.0406) (12288,1.6883) (16384,3.1388) (32768,17.7014)
        };
        \addplot+[mark=o, color=BrickRed] coordinates {
            (512,0.0017) (1024,0.0083) (2048,0.0315) (4096,0.1423) (8192,1.2422) 
            (12288,1.8102) (16384,3.5228) (32768,124.9762)
        };

        \legend{M4 Pro CPU FP16, M4 Pro CPU FP32, M4 Pro GPU FP16, M4 Pro GPU FP32, M4 Pro ANE FP16}
        \end{axis}
    \end{tikzpicture}
    \label{fig:m4multigridTime}
\end{figure}

\begin{figure}[htbp]
    \centering
    \caption{Multigrid mean energy consumption in Mac M1}
    \begin{tikzpicture}
        \begin{axis}[
            width=\linewidth,       
            height=0.55\linewidth,  
            xlabel={$k$ dimension},
            ylabel={Energy consumption (J)},
            grid=both,
            xmin=0, xmax=16384,
            ymin=0, ymax=400,
            ymode=log, 
            tick label style={font=\small},
            label style={font=\small},
            title style={font=\small\bfseries},
            legend cell align={left},
            legend style={
                at={(0.5,-0.33)},
                anchor=north,
                legend columns=2,
                font=\small,
                /tikz/every even column/.append style={column sep=0.4cm}
            },
            scaled ticks=false,
            xtick={0, 2500, 5000, 7500, 10000, 12500, 15000},
            xticklabel style={
                /pgf/number format/fixed,
                /pgf/number format/1000 sep={\,}
            }
        ]

        \addplot+[mark=x, color=cyan] coordinates {
            (512,0.1646) (1024,0.6869) (2048,3.0716) (4096,13.2956)
            (8192,56.7446) (10240,82.4497) (12288,119.6154) (16384,222.4749)
        };
        \addplot+[mark=*, color=RoyalPurple] coordinates {
            (512,0.1388) (1024,0.6430) (2048,3.7206) (4096,18.9411)
            (8192,85.4653) (10240,122.6298) (12288,201.7317) (16384,366.5673)
        };
        \addplot+[mark=square*, color=ForestGreen] coordinates {
            (512,0.0325) (1024,0.0997) (2048,0.4715) (4096,2.0529)
            (8192,8.9146) (10240,14.0475) (12288,20.3608) (16384,36.3901)
        };
        \addplot+[mark=square, color=BurntOrange] coordinates {
            (512,0.0389) (1024,0.1652) (2048,0.6611) (4096,3.1665)
            (8192,13.8963) (10240,21.2623) (12288,30.6568) (16384,56.1054)
        };
        \addplot+[mark=o, color=BrickRed] coordinates {
            (512,0.0190) (1024,0.0874) (2048,0.3614) (4096,1.7174)
            (8192,14.9865) (10240,19.4480)
        };

        \legend{M1 CPU FP16, M1 CPU FP32, M1 GPU FP16, M1 GPU FP32, M1 ANE FP16}
        \end{axis}
    \end{tikzpicture}
    \label{fig:m1multigridConsume}
\end{figure}

\begin{figure}[htbp]
    \centering
    \caption{Multigrid mean energy consumption in Mac M4 Pro}
    \begin{tikzpicture}
        \begin{axis}[
            width=\linewidth,       
            height=0.55\linewidth,  
            xlabel={$k$ dimension},
            ylabel={Energy consumption (J)},
            grid=both,
            xmin=0, xmax=32768,
            ymin=0, ymax=650,
            ymode=log, 
            tick label style={font=\small},
            label style={font=\small},
            title style={font=\small\bfseries},
            legend cell align={left},
            legend style={
                at={(0.5,-0.33)},
                anchor=north,
                legend columns=2,
                font=\small,
                /tikz/every even column/.append style={column sep=0.4cm}
            },
            scaled ticks=false,
            xticklabel style={
                /pgf/number format/fixed,
                /pgf/number format/1000 sep={\,}
            },
            xtick={0, 5000, 10000, 15000, 20000, 25000, 30000}
        ]

        \addplot+[mark=x, color=cyan] coordinates {
            (512,0.0781) (1024,0.2633) (2048,1.0997) (4096,4.2235)
            (8192,20.717) (10240,18.978) (12288,28.558) (16384,79.714)
            (32768,315.548)
        };
        \addplot+[mark=*, color=RoyalPurple] coordinates {
            (512,0.1398) (1024,0.7434) (2048,2.9726) (4096,13.469)
            (8192,43.529) (10240,59.609) (12288,105.628) (16384,167.589)
            (32768,633.660)
        };
        \addplot+[mark=square*, color=ForestGreen] coordinates {
            (512,0.0201) (1024,0.0481) (2048,0.1477) (4096,1.0689)
            (8192,4.622) (10240,8.239) (12288,12.307) (16384,22.699)
            (32768,114.402)
        };
        \addplot+[mark=square, color=BurntOrange] coordinates {
            (512,0.0198) (1024,0.0572) (2048,0.2629) (4096,1.488)
            (8192,7.563) (10240,13.208) (12288,20.252) (16384,37.239)
            (32768,162.884)
        };
        \addplot+[mark=o, color=BrickRed] coordinates {
            (512,0.0150) (1024,0.0446) (2048,0.1810) (4096,0.752)
            (8192,5.323) (12288,8.366) (16384,15.801) (32768,615.773)
        };

        \legend{M4 Pro CPU FP16, M4 Pro CPU FP32, M4 Pro GPU FP16, M4 Pro GPU FP32, M4 Pro ANE FP16}
        \end{axis}
    \end{tikzpicture}
    \label{fig:m4multigridConsumo}
\end{figure}

From the energy consumption side, the ANE presents its strength, although the GPU still demonstrates superior performance. Figures~\ref{fig:m1multigridConsume} and ~\ref{fig:m4multigridConsumo} present the mean power consumption on the M1 and M4 Pro, respectively. Once again, the powerful GPU achieves a better performance rate at the expense of being energy hungry. For small problem sizes (grids smaller than $4096^2$), the ANE exhibits lower energy consumption in terms of Joules, which can be 15\% to 50\% lower than the other devices. For instance, for a grid size of $4096^2$, the ANE on the M4 Pro consumes 0.5 Joules compared to the CPUs 13.47 Joules and the GPU's 1.49 Joules. For larger grid sizes, the ANE's memory limitations prevent its execution, making the GPU the most energy-efficient device.

\subsection{Discussion}

The results presented in Table~\ref{table:speedup_m1_m4}, which uses the M1 CPU's execution time as the baseline, allow us to quantify the inter-generational performance increase between the M1 and M4 Pro chips for each processing unit. This comparison reveals a significant disparity in the scaling improvement: while the CPU and GPU experienced substantial acceleration, the ANE improved by a notably smaller factor.

The general-purpose units (CPU and GPU) demonstrated robust scaling: GPU performance improved by a factor of $3\times$ to $5\times$. For instance, in the Multigrid (8192) benchmark, the acceleration over the M1 CPU rose from $6.57\times$ (M1 GPU) to $25.75\times$ (M4 Pro GPU), representing a generational improvement factor of $3.92\times$; CPU performance also experienced considerable acceleration, reaching up to $4.5\times$. The Multigrid ($12288$) benchmark on the M4 Pro CPU showed an acceleration of $4.54\times$ relative to the M1 CPU.

In contrast, the ANE, despite often remaining the fastest unit in absolute terms, exhibited the lowest inter-generational improvement factor: AI Tasks (YOLOv11l) increased from $19.36\times$ (M1) to $28.96\times$ (M4 Pro). This translates to a generational improvement factor of only $1.50\times$. In Jacobi ($8192$), the ANE's acceleration moved from $3.79\times$ (M1) to $7.34\times$ (M4 Pro), yielding an improvement factor of $1.94\times$. The highest factor for the ANE was observed in Multigrid ($4096$), with an improvement of $2.41\times$.

In summary, while the CPU and GPU experienced performance increases ranging between $3\times$ and $4.5\times$, the ANE's improvement was limited to a range of $1.5\times$ to $2.5\times$. This unequal scaling suggests that the M4 Pro has significantly closed the performance gap that the ANE enjoyed over the CPU and GPU in the M1 generation.

\begin{table}[htbp]
\centering
\caption{Speedup from CPU M1 FP16}
\begin{tabularx}{\linewidth}{l l *{5}{>{\centering\arraybackslash}X}}
\toprule
\textbf{Benchmark} & \textbf{Size} & \textbf{GPU M1} & \textbf{ANE M1} & \textbf{CPU M4Pro} & \textbf{GPU M4Pro} & \textbf{ANE M4Pro}\\
\midrule

\multirow{1}{*}{YOLOv11l}
 & \multirow{1}{*}{\makecell{-}} & 5.43 & 19.36 & 6.82 & 16.36 & 28.96 \\
\midrule

\multirow{2}{*}{GEMM}
 & \multirow{1}{*}{4096} & 1.38 & 2.70 & 1.49 & 3.31 & 3.39 \\
\cmidrule(lr){2-7}
 & \multirow{1}{*}{8192} & 1.83 & \makecell{-} & 1.36 & 4.87 & 3.74 \\
\midrule

\multirow{3}{*}{Jacobi}
 & \multirow{1}{*}{4096} & 8.57 & 8.90 & 3.38 & 24.31 & 16.17 \\
\cmidrule(lr){2-7}
 & \multirow{1}{*}{8192} & 9.47 & 3.79 & 2.49 & 23.42 & 7.34 \\
\cmidrule(lr){2-7}
 & \multirow{1}{*}{12288} & 9.03 & \makecell{-} & 3.93 & 19.20 & 10.60 \\
\midrule

\multirow{3}{*}{Multigrid}
 & \multirow{1}{*}{4096} & 6.88 & 7.21 & 3.96 & 26.54 & 17.40 \\
\cmidrule(lr){2-7}
 & \multirow{1}{*}{8192} & 6.57 & 3.45 & 2.87 & 25.75 & 8.24 \\
\cmidrule(lr){2-7}
 & \multirow{1}{*}{12288} & 6.67 & \makecell{-} & 4.54 & 21.57 & 11.95 \\
\bottomrule
\end{tabularx}
\label{table:speedup_m1_m4}
\end{table}

\section{Conclusions}
\label{conclusions}

This research evaluated the performance and energy efficiency of the Apple Silicon M1 and M4 Pro architectures across a combination of AI and general-purpose computing workloads. The findings confirm the effectiveness of the ANE for accelerating machine learning tasks, establishing it as the leading device for inference and the most efficient in terms of power consumption for these workloads. 

This research also analyzes the behavior of general-purpose workloads (GEMM, Jacobi, and Multigrid) on Apple Silicon. On one hand, the ANE achieves strong performance in GEMM, reaching up to 2.5 TFLOPS on the M1 chip and up to 3.8 TFLOPS on the M4 Pro. This performance is impressive, as it hits 80\% of the GPU's peak performance while using 80\% less power than the GPU on M4 systems. On the other hand, the ANE shows excellent energy efficiency on more complex general-purpose workloads as Jacobi, reaching an impressive efficiency of 49.8 GFlops/W. However, it is important to note that ANE has notable memory limitations that cause runtime errors for large amounts of data.

Focusing on the inter-generational evolution (from the M1 to the M4 Pro), a disparate performance improvement is observed among devices: while the CPU and GPU have experienced significant acceleration (on average $4.9\times$), the ANE has improved significantly less (only $2.3\times$). This suggests a closing performance gap due to substantial baseline improvements in the CPU and GPU. Ultimately, the ANE proves to be a competent and highly efficient device for general-purpose algorithms, particularly those involving small data processing.

As future work, we propose dividing the Multigrid method into several models to run on different devices. The idea is motivated because approximately 50\% of the time is spent on the finest grid (the largest one), and the ANE has shown better performance with smaller grid sizes, so it is convenient to divide the algorithm between the GPU and the ANE, assigning parts according to their optimal performance.

\section*{Acknowledgment}
This paper has been supported by the 
Spanish MINECO under grants PID2021-126576NB-I00 and PID2024-158311NB-I00 funded by MCIN/AEI/10.13039/501100011033 and by FEDER, EU.

\bibliographystyle{IEEEtran}
\bibliography{biblio}

@article{9739030,
  author={Bavikadi, Sathwika and Dhavlle, Abhijitt and Ganguly, Amlan and Haridass, Anand and Hendy, Hagar and Merkel, Cory and Reddi, Vijay Janapa and Sutradhar, Purab Ranjan and Joseph, Arun and Pudukotai Dinakarrao, Sai Manoj},
  journal={IEEE Design \& Test}, 
  title={A Survey on Machine Learning Accelerators and Evolutionary Hardware Platforms}, 
  year={2022},
  volume={39},
  number={3},
  pages={91-116},
  doi={10.1109/MDAT.2022.3161126}
}

@misc{reed2022reinventinghighperformancecomputing,
      title={Reinventing High Performance Computing: Challenges and Opportunities}, 
      author={Daniel Reed and Dennis Gannon and Jack Dongarra},
      year={2022},
      eprint={2203.02544},
      archivePrefix={arXiv},
      primaryClass={cs.DC},
      url={https://arxiv.org/abs/2203.02544}, 
}

@misc{vaswani2023attentionneed,
      title={Attention Is All You Need}, 
      author={Ashish Vaswani and Noam Shazeer and Niki Parmar and Jakob Uszkoreit and Llion Jones and Aidan N. Gomez and Lukasz Kaiser and Illia Polosukhin},
      year={2023},
      eprint={1706.03762},
      archivePrefix={arXiv},
      primaryClass={cs.CL},
      url={https://arxiv.org/abs/1706.03762}, 
}

@inproceedings{6493641,
  author={Esmaeilzadeh, Hadi and Sampson, Adrian and Ceze, Luis and Burger, Doug},
  booktitle={2012 45th Annual IEEE/ACM International Symposium on Microarchitecture}, 
  title={Neural Acceleration for General-Purpose Approximate Programs}, 
  year={2012},
  volume={},
  number={},
  pages={449-460},
  doi={10.1109/MICRO.2012.48}}

@misc{hubner2025,
      title={Apple vs. Oranges: Evaluating the Apple Silicon M-Series SoCs for HPC Performance and Efficiency}, 
      author={Paul Hübner and Andong Hu and Ivy Peng and Stefano Markidis},
      year={2025},
      eprint={2502.05317},
      archivePrefix={arXiv},
      primaryClass={cs.AR},
      url={https://arxiv.org/abs/2502.05317}, 
}

@article{Opt_for_GPUs,
author = {Hijma, Pieter and Heldens, Stijn and Sclocco, Alessio and van Werkhoven, Ben and Bal, Henri E.},
title = {Optimization Techniques for GPU Programming},
year = {2023},
issue_date = {November 2023},
publisher = {Association for Computing Machinery},
address = {New York, NY, USA},
volume = {55},
number = {11},
issn = {0360-0300},
url = {https://doi.org/10.1145/3570638},
doi = {10.1145/3570638},
journal = {ACM Comput. Surv.},
month = mar,
articleno = {239},
numpages = {81}
}

@INPROCEEDINGS{AI_accelerator,
  author={Reuther, Albert and Michaleas, Peter and Jones, Michael and Gadepally, Vijay and Samsi, Siddharth and Kepner, Jeremy},
  booktitle={2021 IEEE High Performance Extreme Computing Conference (HPEC)}, 
  title={AI Accelerator Survey and Trends}, 
  year={2021},
  volume={},
  number={},
  pages={1-9},
  doi={10.1109/HPEC49654.2021.9622867}}

@inproceedings{Alexnet,
author = {Krizhevsky, Alex and Sutskever, Ilya and Hinton, Geoffrey E.},
title = {ImageNet classification with deep convolutional neural networks},
year = {2012},
publisher = {Curran Associates Inc.},
address = {Red Hook, NY, USA},
booktitle = {Proceedings of the 26th International Conference on Neural Information Processing Systems - Volume 1},
pages = {1097–1105},
numpages = {9},
location = {Lake Tahoe, Nevada},
series = {NIPS'12}
}

@article{AppleM1-MPR,
  title={Apple ships its first PC processor},
  author={Jani Aakash},
  journal={Microprocessor Report},
  year={2021},
  publisher={Linley Group},
}

@article{DL_HW_accelerator,
author = {Silvano, Cristina and Ielmini, Daniele and Ferrandi, Fabrizio and Fiorin, Leandro and Curzel, Serena and Benini, Luca and Conti, Francesco and Garofalo, Angelo and Zambelli, Cristian and Calore, Enrico and Schifano, Sebastiano and Palesi, Maurizio and Ascia, Giuseppe and Patti, Davide and Petra, Nicola and De Caro, Davide and Lavagno, Luciano and Urso, Teodoro and Cardellini, Valeria and Cardarilli, Gian Carlo and Birke, Robert and Perri, Stefania},
title = {A Survey on Deep Learning Hardware Accelerators for Heterogeneous HPC Platforms},
year = {2025},
issue_date = {November 2025},
publisher = {Association for Computing Machinery},
address = {New York, NY, USA},
volume = {57},
number = {11},
issn = {0360-0300},
url = {https://doi.org/10.1145/3729215},
doi = {10.1145/3729215},
journal = {ACM Comput. Surv.},
month = jun,
articleno = {286},
numpages = {39}
}

@MastersThesis{Jonathan_thesis2025,
  title     = "Performance Analysis of the Apple AMX Matrix Accelerator",
  author    = "Zhou, Jonathan",
  month     = "Aug",
  year      = "2025",
  url       = "https://commit.csail.mit.edu/papers/2025/Jonathan_Zhou_SB_Thesis.pdf",
  type      = "S.B. Thesis",
  address   = "Cambridge, MA",
  school    = "Massachusetts Institute of Technology"
}

\end{document}